\documentclass[epj]{svjour}
\usepackage{graphics}
\usepackage{psfig}

\newcommand{\bey}[1]{\begin{eqnarray} \label{#1}}
\newcommand{\eey}{\end{eqnarray}}
\newcommand{\beq}[1]{\begin{equation} \label{#1}}
\newcommand{\eeq}{\end{equation}}

\begin{document}

\title{Field Theoretic Approach to Long Range Reactions}

\author{Jeong-Man Park\thanks{Present address: 
              Department of Physics, The Catholic University
              of Korea, Seoul, Korea}
 and Michael W. Deem
}
\institute{Chemical Engineering Department, University of California,
           Los Angeles, CA  90095-1592 USA}
\date{Received: date / Revised version: date}
\abstract{
We analyze bimolecular reactions that proceed
by a long-ranged reactive interaction, using
a field theoretic approach that takes into account
fluctuations.
 We consider both the
one-species, $A+A \to \emptyset$ reaction and the
two-species, $A+B \to \emptyset$ reaction.  We consider both mobile
and immobile reactants, both in the presence and in the
absence of adsorption.
}

\maketitle

\section{Introduction}
\label{introduction}
Most chemical reactions occur through a local mechanism, where
reaction occurs at a finite rate only when reactants are closer than
a capture radius.  In some unusual cases, however,
the reaction can proceed by a long-range mechanism
\cite{Forester}.
Exciton decay by a multipolar interaction is one such example.
In these long-range
cases, the rate of reaction between two reactants is distance dependent:
$w(r) = \gamma / r^n$ for two reactants separated by a distance $r$.
Here the strength of the interaction is given by
$\gamma = R_0^n/\tau$, where $R_0$ is the F{\"o}rster radius, and
$\tau$ is the decay lifetime.
A variety of electronic energy transfer reactions can be modeled
by this form \cite{Klafter2}.  More generally, an exponential form
of the reaction rate may be considered, as in
electron-hole recombination due to wave function overlap or
barrier tunneling.  The exact form of the interaction can be
quite complex \cite{Klafter3}, and so even these shorter-range
cases may equally well be modeled by
the multipolar form, \emph{e.g.}\ with $n=12$ \cite{Silby,Klafter1}.
A wide variety of physical systems exhibit
such long-range reaction, including
biological electron transport, fluorescent decay of electron-hole
pairs in amorphous semiconductors, scavenging of trapped electrons
in organic glasses, and decay of localized electronic states is
mixed organic solids.

The case of long-range
reaction has received less theoretical
attention than has the more common case of local reaction. 
 The one-species, $A+A \to \emptyset$
reaction between immobile reactants without adsorption
has been investigated by placing bounds on a 
mean-field treatment \cite{Burlatsky2}.  An interesting power-law
decay of the concentration was found in $d$ spatial
dimensions at long times:
$c(t) \sim a t^{-d/n}$  as $t \to \infty$ for $n>d$.
An extension to the $A+B \to \emptyset$
case gave the scaling 
$c(t) \sim a t^{-d/(2 n-d)}$  as $t \to \infty$ for $n>d$.
 The single-species reaction, $A +A \to \emptyset$
reaction between immobile reactants in the presence of adsorption
has been investigated with a similar
treatment \cite{Oshanin2}.  
Adsorption corresponds to the creation of reactants, $\emptyset \to A$, with
the conventional adsorption rate $J$.
A non-trivial scaling of the concentration
with the adsorption rate was found:
$c(J) \sim a J^{d/(n+d)}$ as $J \to 0$, $n>d$.  In principle, however,
 these mean-field results could be modified by renormalization effects
in low spatial dimensions.

In this paper, we apply the 
renormalization group approach to reaction kinetics to derive the
asymptotically exact behavior of these long range reactions.
There are eight distinct physical cases, 
considering all possibilities of one- or two-species reactions,
with or without adsorption, and mobile or immobile reactants.
Our paper is organized as follows.
In section \ref{sec0}, we review the field theoretic approach,
displaying the actions appropriate for both the one- and two-species
reactions in the general case.  In section \ref{sec1}, we
derive the behavior of mobile reactants in the long-time limit
in the absence of adsorption.  In section \ref{sec2},  we
derive the long-time behavior of immobile reactants 
in the absence of adsorption.  
In section \ref{sec3}, we derive the steady-state behavior of
mobile reactants in the limit of small adsorption rates.
Finally, in section \ref{sec4} we derive the steady-state behavior of
immobile reactants in the limit of small adsorption rates.
We summarize our results in section \ref{conclusion}.

\section{Field Theoretic Formulation of Reaction Dynamics}
\label{sec0}

We consider the general case of a bimolecular reaction
in $d$ spatial dimensions.  The reaction occurs as a result of a long-range
interaction.  For convenience, we 
define the reaction to proceed on a cubic lattice.
We use a continuous-time master equation to define
how the probability of any given configuration of reactants 
changes with time.  The master equation for the
$A + A \to \emptyset$ reaction is
\bey{1}
&&\frac{\partial P(\{n_i\}, t)}{\partial t} = 
\nonumber \\ &&
\frac{D}{h^2} \sum_{ij}
[ (n_j +1)P(\ldots, n_i -1,n_j +1, \ldots, t)
- n_i P]
\nonumber \\ &&
+ \frac{1}{2} \sum_{ik} w_{ik} [(n_i+1)(n_k +1)
 P (\ldots, n_i+1, n_k+1, \ldots ,t)
\nonumber \\ &&
~~~~~~~~~~~~~~~-n_i n_k P ] 
\nonumber \\ &&
+ J h^d \sum_i [P(\ldots, n_i-1, \ldots) - P]
\ ,
\eey
where $h$ is the lattice spacing,
$J$ is the rate of adsorption, 
D is the diffusivity, 
$n_i$ is the number of $A$ species on lattice site $i$,
and $w_{ik}$ the distance-dependent reactive interaction.
The summations over
$i$ and $k$ are over all sites on the lattice, and the summation over $j$ is
over all nearest neighbors of site $i$.  For simplicity, we
choose to place the species initially
at random on the lattice, with average number density $n_0$.

Using the coherent state representation, we map this master 
equation onto a field theory \cite{Peliti,Lee1,Deem1}.  We
find that the reactant concentration, averaged over the random
initial conditions and the random statistics of the reaction-diffusion
process, is
\beq{2}
c_A(\vec{x}, t) = \langle a(\vec{x}, t) \rangle \ ,
\eeq
where the average is taken with respect to $\exp(-S_{AA})$.  The
action is given by
\bey{3}
&&S_{AA} = \int d^d \vec{x}\, \int_0^{t_{\rm f}} d t\,
 \bar a(\vec{x}, t) \left[
\partial_t - D \nabla^2 + \delta(t)
 \right]
 a(\vec{x}, t)
 \nonumber \\ 
&&  -n_0 \int d^d \vec{x}\, \bar a(\vec{x}, 0) 
  -J \int d^d \vec{x}\, \int_0^{t_{\rm f}} d t\, \bar a(\vec{x}, t) 
 \nonumber \\ 
&& + \frac{1}{2} \int d^d \vec{x}\, d^d \vec{y}\, \int_0^{t_{\rm f}} d t\,
  \bar a(\vec{x}, t) a(\vec{x}, t) w(\vert \vec{x} - \vec{y} \vert )
 \bar a(\vec{y}, t) a(\vec{y}, t) 
 \nonumber \\ 
&&  + \int d^d \vec{x}\, d^d \vec{y}\, \int_0^{t_{\rm f}} d t\,
  \bar a(\vec{x}, t) a(\vec{x}, t) w(\vert \vec{x} - \vec{y} \vert )
  a(\vec{y}, t) \ .
\eey
The time $t_{\rm f}$  must be larger than the longest time for
which we wish to make predictions.
Special cases of this general formulation are given as limiting
values.  For immobile reactants, for example, we set $D = 0$.
If there is no adsorption, we set $J = 0$.

For dissimilar reactants, the formulation is slightly different.
The master equation is given by
\bey{4}
&&\frac{\partial P(\{ m_i \}, \{ n_i \}, t) }{\partial t}  =
\nonumber \\
&&\frac{D}{h^2} \sum_{ij}
[(m_j +1)P(\ldots, m_i -1, m_j +1, \ldots, t)
-  m_i P]
\nonumber \\
&&+ \frac{D}{h^2} \sum_{ij}
[ (n_j +1)P(\ldots, n_i -1,n_j +1, \ldots, t)
-  n_i P]
\nonumber \\
&&+ \sum_{ik} w_{ik} [(m_i+1)(n_k +1)
 P (\ldots, m_i+1, n_k+1, \ldots ,t)
\nonumber \\
&&~~~~~~~~~~~~~~~ -m_i n_k P ]
\nonumber \\
&&+ J h^d \sum_i [P(\ldots, m_i-1, \ldots) - P]
\nonumber \\
&&+ J h^d \sum_i [P(\ldots, n_i-1, \ldots) - P]
 \ ,
\eey
where now
$m_i$ and $n_i$ are the number of $A$ and $B$
species, respectively, on lattice site $i$.  We have assumed for
simplicity that the diffusion constants of the two species are
the same.  So as to reach an interesting scaling regime, we have
set the adsorption rates of the two species to be the same.
Finally, we will set the average initial concentrations of the
two reactants to be the same.

We derive the  field theory from this master equation using the
coherent state representation \cite{Lee2,Deem2}.
The averaged concentrations are given by averages over
two distinct fields:
\bey{5}
c_A(\vec{x}, t) &=& \langle a(\vec{x}, t) \rangle \nonumber \\
c_B(\vec{x}, t) &=& \langle b(\vec{x}, t) \rangle \ ,
\eey
where now the average is with respect to $\exp(-S_{AB})$.
The action for dissimilar species is given by 
\bey{6}
&&S_{AB} =
\nonumber \\
&&\int d^d \vec{x}\, \int_0^{t_{\rm f}} d t\,
 \bar a(\vec{x}, t) \left[
\partial_t - D \nabla^2 + \delta(t)
 \right]
 a(\vec{x}, t)
 \nonumber \\ 
&&+ \int d^d \vec{x}\, \int_0^{t_{\rm f}} d t\,
 \bar b(\vec{x}, t) \left[
\partial_t - D \nabla^2 + \delta(t)
 \right]
 b(\vec{x}, t)
 \nonumber \\ 
&&  -n_0 \int d^d 
\vec{x}\,  \left[ \bar a(\vec{x}, 0) + \bar b(\vec{x}, 0 \right]
 \nonumber \\ 
&&  -J \int d^d \vec{x}\, \int_0^{t_{\rm f}} d t\,
     \left[ \bar a(\vec{x}, t) +\bar b(\vec{x}, t) \right]
 \nonumber \\ 
&& + \int d^d \vec{x}\, d^d \vec{y}\, \int_0^{t_{\rm f}} d t\,
  \bar a(\vec{x}, t) a(\vec{x}, t) w(\vert \vec{x} - \vec{y} \vert )
  \bar b(\vec{y}, t) b(\vec{y}, t) 
 \nonumber \\ 
&&  + \int d^d \vec{x}\, d^d \vec{y}\, \int_0^{t_{\rm f}} d t\,
  \bar a(\vec{x}, t) a(\vec{x}, t) w(\vert \vec{x} - \vec{y} \vert )
  b(\vec{y}, t)
 \nonumber \\ 
&&  + \int d^d \vec{x}\, d^d \vec{y}\, \int_0^{t_{\rm f}} d t\,
  \bar b(\vec{x}, t) b(\vec{x}, t) w(\vert \vec{x} - \vec{y} \vert )
  a(\vec{y}, t) \ .
\eey

\begin{figure}[t]
\centering
\leavevmode
\psfig{file=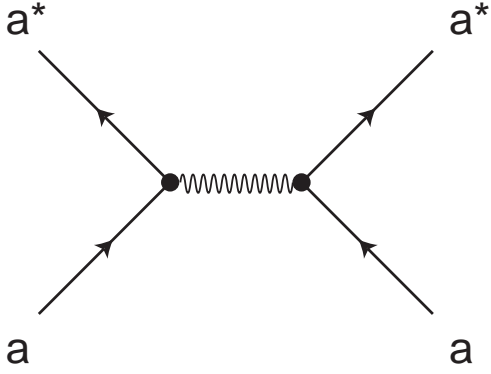,height=2in}
\caption[]
{\label{fig1}
The vertex in the field theory of the $A+A \to 0$ reaction.
Time increases in the direction of the arrows.
The wavy line indicates reaction with the
distance-dependent rate $w(r)$.
}
\end{figure}

\begin{figure}[t]
\centering
\leavevmode
\psfig{file=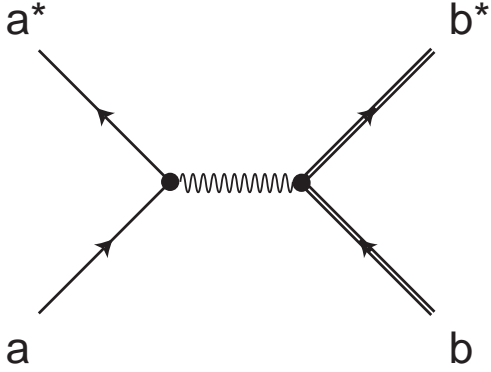,height=2in}
\caption[]
{\label{fig2}
The vertex in the field theory of the $A+B \to 0$ reaction.
}
\end{figure}

The long-ranged interaction shows up in the non-quadratic terms
of this field theory.  The vertex that we must consider for the
single-species reaction is shown in figure \ref{fig1}.
Similarly, the vertex we must consider in the two-species
reaction is shown in figure \ref{fig2}.
We will use both a regularized interaction
\beq{6a}
w_R(r) = \gamma / (r^2 + R^2)^{n/2}
\eeq
and a pure power-law form of the interaction.
The Fourier transform of the interaction is given by
\bey{7}
\hat w_R(\vec{k}) &=& \int d^d \vec{r}\, \exp(i \vec{k} \cdot \vec{r})
w_R(\vec{r})
\nonumber \\
&= &
\frac{\gamma \pi^{d/2} (k R/2)^{(n-d)/2}}
{R^{n-d} \Gamma(n/2)}
2 K_{(n-d)/2}( k R) \ ,
\eey
where $\Gamma(x)$ is the Gamma function, and $K_l(x)$ is the
modified Bessel function of the second kind.
At long-times, for low adsorption rates, the reactants tend to be
widely separated, and the regularization does not affect the
reaction kinetics.  In the absence of regularization, we find
\bey{8}
\lim_{R \to 0} [ \hat w_R(0) - \hat w_R(k)] &=&
\hat w(0) - \hat w(k) 
\nonumber \\ &= &
\frac{\gamma \pi^{d/2} \Gamma[(d-n)/2]}{2^{n-d} \Gamma(n/2)} k^{n-d} \ .
\eey
This equation is well-defined by the Fourier integral for
$d < n < d+2$.  It is derived for all other $n$
by analytic continuation from the relation
$FT\{ \nabla^2 w \} = -k^2 FT\{ w \}$.

\section{Mobile Reactants without Adsorption}
\label{sec1}

In this section we address the simplest case of bimolecular
reactions between mobile reactants in the absence of adsorption.
That is, we set $J=0$ in the field-theoretic actions.
We use renormalization group theory to calculate the 
asymptotic reactant concentrations in the long-time limit.

\subsection{The $A+A \to \emptyset$ Reaction}
\label{sec1a}
\begin{figure}[t]
\centering
\leavevmode
\psfig{file=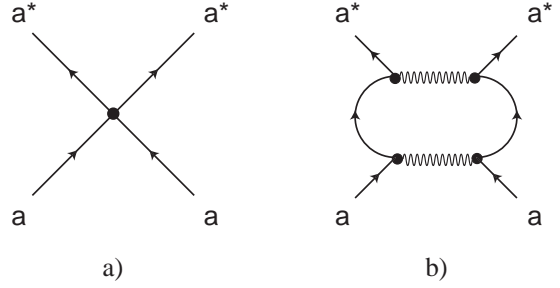,height=1.5in}
\caption[]
{\label{fig3}
The local vertex and vertex correction
in the field theory of the $A+A \to 0$ reaction.
a) The effective local interaction, into which the non-local
vertex flows.  b) The
vanishing correction to the non-local interaction.
}
\end{figure}

For the single species reaction, we find the flow equations to be
\bey{9}
\frac{d \ln n_0}{d l} &=& d
\nonumber \\
\frac{d \ln D}{d l} &=& z-2
\nonumber \\
\frac{d \ln \gamma}{d l} &=& z-n \ ,
\eey
where $z$ is the dynamical exponent.  These flow equations are
exact to all orders.  
Time renormalizes as $t(l) = e^{-z l} t$.
To reach the fixed point,
which requires the propagator to reach a fixed point as well, we
set $z=2$.  
  We see that the strength of the
long-range reaction, $\gamma$, 
appears to be irrelevant when $n>2$.  What this result actually implies
is that the long-range nature of the reaction is irrelevant.  Certainly,
though, the reaction event itself is relevant.  We can, therefore,
use the effective interaction
$w(r) = \lambda \delta (\vec{r})$.  This replacement is
possible only when the interaction is integrable, $n > d$, which
we always require.  For non-integrable interactions, the
reactant concentration decays to zero immediately.
The scaling for the effective reaction rate is
\beq{10}
\frac{d \ln \lambda}{d l} = z-d - \frac{\lambda}{4 \pi D} \ .
\eeq
This flow equation is a one-loop expansion for small $\lambda$.
The flow equations \ref{9} and \ref{10}
 are derived by considering the vertex corrections
shown in figure \ref{fig3}.

The renormalization group transformation relates the original system
at long times and low concentrations to another, renormalized
system at short times and high concentrations.
In fact, we integrate the flow equations until the renormalized time is
short enough so that we can use simple mean field theory \cite{Deem1}.
The matching time 
\beq{11}
t(l^*) = t e^{-\int_0^{l^*} z(l) d l } = t_0
\eeq
is chosen to be on the order of $h^2/ (2 D)$.
Calculating the renormalized
reactant concentration at short times with
 mean field theory, we find
\bey{12}
c[t(l^*); l^*] &=& \frac{1}{1/n_0(l^*) + \lambda(l^*) t(l^*)}
\nonumber \\
&\sim& \frac{1}{\lambda(l^*) t(l^*)} ~\mathrm{as}~ l^* \to \infty.
\eey
The concentration of the original system is related to that of the  effective,
renormalized system by scaling:
\beq{13}
c(t) = e^{-d l} c[t(l); l] \ .
\eeq
Combining equations \ref{12} and \ref{13}, and using equation \ref{11} to
express the result in terms of $t$ rather than $l^*$, we find
\beq{14}
c_A(t) \sim \left\{
\begin{array}{ll}
\frac{1}{\lambda t}& , d > 2\\[0.1in]
\frac{\ln (t/t_0)}{8 \pi D t}& ,  d = 2\\[0.1in]
\frac{1}{\lambda^* t_0} \left( \frac{t}{t_0} \right)^{-d/2}& , d < 2
\end{array}
\right.    ~\mathrm{as}~ t \to \infty.
\eeq
Here, we have used the fact that $\lambda(l)$ goes to a fixed point
value for $d < 2$. To first order in $2 - d$, we find from
equation \ref{10} that $\lambda^* = 4 \pi D (2 - d)$.

The predicted decay in  equation \ref{14} comes from assuming that the 
reaction will be diffusion limited at long times.  This is generally
true, but the results of section \ref{sec2a} show that when $d \le 2$,
the long-range nature of the reaction is relevant
when $n \le 2$.  Our prediction \ref{14}, therefore, is limited to the cases
of $d > 2$ or $d \le 2$ and $n > 2$.

\subsection{The $A+B \to \emptyset$ Reaction}
\label{sec1b}

\begin{figure}[t]
\centering
\leavevmode
\psfig{file=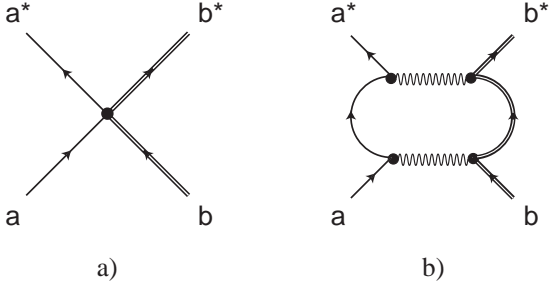,height=1.5in}
\caption[]
{\label{fig4}
The vertex corrections
in the field theory of the $A+B \to 0$ reaction.
a) The effective local interaction, into which the non-local
vertex flows.  b) The
vanishing correction to the non-local interaction.
}
\end{figure}

For the two species reaction, we find the same flow equations as
in equations \ref{9} and \ref{10}.  The vertex corrections
that lead to this result are shown in figure \ref{fig4}.
As with the single species
reaction, the long-range nature is irrelevant when $n>2$.  
At long times, the behavior of the long-range case 
is identical to that of the local case.
This implies, as before, that $z=2$.
  The matching is somewhat more complicated than
that of the single species reaction, due to the segregation between
the $A$ and $B$ species that occurs.
 The result is \cite{Deem2}
\beq{15}
c_A(t) = c_B(t) \sim \left\{
\begin{array}{ll}
\frac{1}{\lambda t} & , d > 4\\[0.1in]
\frac{\sqrt n_0}{ \sqrt \pi (8 \pi D t)^{d/4} } & , 1 \le d < 4
\end{array}
\right.    ~\mathrm{as}~ t \to \infty.
\eeq
As in section \ref{sec1a}, the prediction in equation
\ref{15} is limited to the diffusion-limited regime, which
section \ref{sec2b} shows 
occurs for $d > 4$ or $d \le 4$ and $ n > 4$.

In summary, for mobile reactants, the long-range nature of the
reaction is irrelevant in the diffusion-limited regime.
The long-time scaling of the concentration is the same as that of
a reaction with a
local interaction for both the one- and two-species reactions.

\section{Immobile Reactants without Adsorption}
\label{sec2}

The case of immobile reactants in the absence of
adsorption, $D=0$ and $J=0$, is rather different.
This is because for immobile reactants, the long-range 
and short-range cases are quite distinct.
This is simply because a short-range reaction
stops as soon as no more than one reactant
occupies each lattice site.  The long-range reaction,
on the other hand, continues until either zero or one reactant
remains in the entire system.  As we will see, this 
essential difference leads to a more complicated matching
procedure.

\subsection{The $A+A \to \emptyset$ Reaction}
\label{sec2a}

As before, we use renormalization group theory to analyze
the field theory in the
long-time regime. Now, however, we must be careful with the
definition of the interaction.  We use the regularized
version in equation \ref{6a}.  Since we are interested in the
long-time regime, where the particle density is low and the
particles are widely separated, we expand the interaction
for large $r$:
\beq{16}
w_R(r)  = \frac{\gamma_0}{r^n} + \frac{\gamma_2 }{r^{n+2}} + \ldots ,
\eeq
where $\gamma_0 \equiv \gamma$.
We find the flow equations  to be
\bey{16a}
\frac{d \ln n_0}{d l} &=& d
\nonumber \\
\frac{d \ln \gamma_i}{d l} &=& z-n-i \ .
\eey
We immediately see that the higher order terms in the reaction
rate, $\gamma_i$ for $i > 0$, are less relevant than $\gamma_0$.
So as to reach a fixed point, we set $z=n$.  
The flow equations for $n_0$ and $\gamma_0$ are then asymptotically
exact.

While the regularization of the interaction $w(r)$ is irrelevant for
long-times, it is important in the matching limit.
Indeed, we integrate the
flow equations until the density is on the order of 
$c[t(l^*); l^*] \approx 1 / R^d$.  In that limit, we find
\beq{17}
c(t_0; l^*) = \frac{1}{n_0^{-1}(l^*) + \hat w(0) t_0} \sim
 \frac{1}{\hat w(0) t_0} ~\mathrm{as}~ l^* \to \infty\ ,
\eeq
where $\hat w(0) = (\mathrm{const}) \gamma R^{d-n}$ is not a function of $l$.
We then find
\beq{18}
c(t)  \sim \left\{
\begin{array}{ll}
\frac{1}{\hat w(0) t_0} \left( \frac{t}{t_0} \right) ^{-d/n} & , n > d\\[0.1in]
0 & , n \le d
\end{array}
\right.  ~\mathrm{as}~ t \to \infty.
\eeq
As indicated, the specific value of the cutoff affects the
prefactor, but not the exponent, in the long-time scaling.
When the interaction is not integrable, $n \le d$, the concentration
immediately decays to zero in the limit of a large system size.

While our treatment of this case has required use of a cutoff,
the reaction process defined by the
master equation is well-defined even in the
absence of a cutoff.  This is because any reactants experiencing the
infinite reaction rate $w(0)$ simply react immediately at the onset.
Indeed, in the limit $R \to 0$ and $h \to 0$,
there is an exact scaling relation that relates the system at
long times to a renormalized system at short times.
  If we rescale 
time as $t^* = \gamma n_0^{n/d} t$ and space as
$\vec{x}^* = n_0^{1/d} \vec{x}$, we find that
\beq{19}
c(\vec{x}, t) = n_0 c^*(\vec{x}^*, t^*) \ ,
\eeq 
where $c^*$ is the exact solution of the master equation \ref{1} for
$n_0 = 1$ and $\gamma=1$.
This scaling relation explicitly demonstrates that
a cutoff is not required.  Indeed, this scaling relation
is exactly what the renormalization group treatment is intended
to derive.  Essentially, this problem has so few time and length
scales that the renormalization group approach does not provide
any additional simplification.  

On a deeper level, the scaling 
relation \ref{19} implies that there is no non-trivial renormalization of
the effective interaction for this problem.  Mean field theory,
therefore, will be accurate for both long times and short times.
If we assume that there are no correlations among the
reactants, we can write
\beq{20}
\frac{d c}{d t} = - c \int_R^\infty d r\, S_d r^{d-1} \frac{\gamma}{r^n}
c(r) \ ,
\eeq
where $S_d = 2 \pi^{d/2} / \Gamma(d/2)$ is the surface area in
$d$ dimensions.
Integrating this equation using the approximation that there is
only one reactant per correlated region,
$c(t) \approx d / [S_d R^d(t)]$, we find
\beq{21}
c(t) \sim \frac{d}{S_d} \left( \frac{\gamma n t}{n-d} \right)^{-d/n}
   ~\mathrm{as}~ t \to \infty,~~~ n>d.
\eeq
This relation gives the same power-law decay as does equation \ref{18}.
Moreover, it satisfies the exact scaling relation \ref{19}.
This approximate approach has essentially assumed that the radial
correlation function rises from zero to unity at the average
distance of interparticle separation.  A more careful treatment of
the correlation function
would likely lead to a more accurate prefactor for the asymptotic
concentration decay, without changing the exponent \cite{Kopelman}.

\subsection{The $A+B \to \emptyset$ Reaction}
\label{sec2b}
For the two-species reaction, we find the
same flow equations as for the single-species reaction,
equation \ref{16a}.  The matching is different, however, because
the $A$ and $B$ reactants segregate in the long-time limit, just
as they do in the mobile case.  The segregation of the reactants 
requires us to adopt a different dynamical exponent.
We argue that the degree of reaction between 
a typical reactant and a patch of reactants of volume $\Delta V$
separated by a distance  $r$ is invariant at the fixed point.
The typical number of reactants in a volume $\Delta V$
is proportional  to $(\Delta V)^{1/2}$ in a segregated system,  and so
we require $\gamma (\Delta V)^{1/2}  \Delta t / r^n$ to be invariant
under the spatial and temporal rescaling in the renormalization
group procedure.  This condition gives $z  = n-d/2$.

To analyze the matching regime, we use the mean field equations
\bey{22}
\frac{\partial c_A}{\partial t} &= &
- c_A(\vec{x},t) \int d^d \vec{y}\, c_B(\vec{y}, t) w(\vert\vec{x} - \vec{y}\vert)
\nonumber \\
\frac{\partial c_B}{\partial t} &= &
- c_B(\vec{x},t) \int d^d \vec{y}\, c_A(\vec{y}, t) w(\vert\vec{x} - \vec{y}\vert)
\ .
\eey
Since we expect segregation, we rewrite these equations in the
variables $\phi = (c_A - c_B)/2$ and $\rho = (c_A + c_B)/2$.
We make the approximation of uniform total density: $\rho(\vec{x}, t) \approx
\rho(t)$.  This approximation is valid in the limit of weak
segregation, $\phi \ll \rho$, and in the limit of
strong segregation, $\rho = \vert \phi \vert$, except near domain
boundaries.  With this assumption, we find
\beq{23}
\frac{\partial \phi}{\partial t} = \rho (\phi * w)
-\phi \rho \hat w(0)
\eeq
and
\beq{23a}
\frac{\partial \rho}{\partial t} = \phi (\phi * w) -
\rho^2 \hat w(0)  \ .
\eeq
Equation \ref{23} can be solved, since $\rho$ depends on time only:
\beq{24}
\hat \phi(\vec{k}, t ) = 
\hat \phi(\vec{k}, 0 ) \exp\left\{ -[\hat w(0) - \hat w(k)]
\int_0^t d t' \rho(t') \right\} \ .
\eeq
Taking the average of equation \ref{23a} and using
$\langle \phi(\vec{r}, 0) \phi(\vec{r}', 0) \rangle
= n_0 \delta(\vec{r}  - \vec{r}')/2$, we find
\bey{25}
\frac{d \langle \rho \rangle }{d t} &=& \frac{n_0}{2}
\int_\vec{k} \hat w(\vec{k}) \left\langle \exp\left\{
-2 [\hat w(0) - \hat w(k)]
\int_0^t d t' \rho(t')
\right\} \right\rangle
\nonumber \\
&&- \left\langle \rho^2 \right\rangle \hat w(0) \ ,
\eey
where the notation $\int_\vec{k}$ stands
for $\int d^d \vec{k} / (2 \pi)^d$.
Note that  all the terms in equation \ref{25} scale as
$e^{d l^* /2}$ in the matching limit.  We also know that
the initial reactant concentration decreases from 
$n_0 e^{d l^*}$ to
$\sqrt n_0 e^{d l^*/2}$
in a time of the order of $1/[\hat w(0) n_0 e^{d l^*}]$
\cite{Deem2}.  Knowing this, and using the scaling
relation \ref{13}, 
 we see that the concentrations must scale as
$c_A = c_B \sim (\mathrm{const}) t^{-d / (2 n -d)}$ in the long-time
limit.

The exact scaling relation \ref{19} also applies in this case, however, and so
there are no non-mean-field effects introduced by
renormalization.
Knowing this, we can analyze the approximate mean-field theory
\ref{25} more thoroughly to obtain an estimate of the prefactor in
the long-time decay.  We perform the calculation with
the non-regularized interaction of equation \ref{8}.
 First, we calculate the average fluctuations of $\phi$:
\beq{26}
\langle \phi^2 \rangle = \frac{n_0 S_d}{2 \zeta} \frac{\Gamma(d/\zeta)}
{\left[ 2 \alpha \int_0^t d t' \rho(t') \right]^{d/\zeta} } \ ,
\eeq
where we have defined $\zeta = n-d$ and
\beq{27}
\alpha = \frac{\gamma \pi^{d/2} \Gamma(-\zeta/2) }{2^\zeta \Gamma(n/2)} \ .
\eeq
Balancing the terms in equation \ref{23a}, we find that the
two right-hand terms must be equal,
$\rho(t) \sim \vert \phi(t) \vert$.
Indeed, setting $\phi = a t^{-\delta}$, we find
$\delta = d / (2n-d)$.  We also find an explicit form
for $a$ from equation \ref{26} by noting that since $\phi$ is a
Gaussian field,
$\langle \vert \phi(t) \vert \rangle = 
[2 \langle  \phi^2 \rangle / \pi]^{1/2}$.
We finally obtain
\bey{28}
\langle \rho \rangle &\sim& n_0^{(n-d)/(2 n-d)} \alpha ^{-d/(2 n-d)}
\left[ \frac{S_d \Gamma(d/\zeta)}{2 \zeta}\right]^{(n-d)/(2 n -d)}
\nonumber \\
&&\times \left( \frac{\zeta}{2 n -d} \right)^{d/(2 n-d)}
\left(\frac{2}{\pi}\right)^{1/2}
t^{-d/(2 n -d)}
 \nonumber \\&&
\mathrm{as}~ t \to \infty,~~~ n>d \ .
\eey
This exponent is in agreement with the scaling argument
 of Burlatsky and Chernoutsas
\cite{Burlatsky2}.
This result can be non-dimensionalized exactly as predicted by
equation \ref{19}.  Our approach here 
has made use of a self-consistent approximation.
In particular, we integrated equation \ref{23} by assuming that
$\rho(t)$ was a given function.  We then, however, made the
identification $\rho = \vert \phi \vert$ to obtain the prefactor
in equation \ref{28}.

As in the single-species case, we can also take a correlation
function approach to the calculation of the long-time concentration
profile.  We have
\beq{29}
\frac{d c_A}{d t} = - c_A \int_R^\infty d r\, S_d r^{d-1} \frac{\gamma}{r^n}
c_B(r) \ .
\eeq
Again assuming that the radial distribution function rises from zero
to unity at the average interparticle separation $r=R$, we find
$c_A(t) = \{n_0 d/[S_d R^d(t)]\}^{1/2}$ in the segregated limit.
Integrating equation \ref{29}, we find
\bey{30}
c_A(t) &=& c_B(t) \sim n_0 
\left( \frac{2 n  - d}{n - d} \right)^{- d / (2 n - d)}
\left( \frac{d}{S_d} \right)^{n / (2 n - d)}
\nonumber \\
&& \times \left( n_0^{n/d} \gamma t \right)^{-d/(2 n -d)} 
  ~\mathrm{as}~ t \to \infty,~~~ n>d.
\eey
This result satisfies the exact scaling relation \ref{19}.
As in the single-species case, a
 more accurate correlation function approach would likely
lead to a better prefactor, without modification of the
exponent \cite{Kopelman2}.

\section{Mobile Reactants with Adsorption}
\label{sec3}

We now turn to consider the case of
long-range reaction of mobile reactants
in the presence of adsorption.
That is, we consider the general case defined by
the master equations \ref{1} or \ref{4}, with $D \ne 0$ and $J \ne 0$.
To access an interesting scaling regime, we consider
the limit $J \to 0$ and seek to understand how the
concentration scales with $J$ in this limit.

\subsection{The $A+A \to \emptyset$ Reaction}
\label{sec3a}

We find the flow equations
for the single-species reaction to be
\bey{31}
\frac{d \ln n_0}{d l} &=& d
\nonumber \\
\frac{d \ln J}{d l} &=& z+d
\nonumber \\
\frac{d \ln D}{d l} &=& z-2
\nonumber \\
\frac{d \ln \gamma}{d l} &=& z-n \ .
\eey
These flow equations are exact to all orders.
We see that, as for the case of no adsorption,
the long-range nature of the reaction is irrelevant when $n>2$.
Of course, the presence of the reaction is important, so we
set the interaction to the effective value
$w(r)  = \lambda \delta (\vec{r})$.
We require the interaction to be integrable, $n>d$, as always.
The flow equation for $\lambda$ is the same as in equation \ref{10}.
To reach the fixed point, we set $z=2$.
We integrate these flow equations until $J(l^*) = 
e^{(2+d) l^*} J = J_0 \approx D / h^{2+d}$.
For this renormalized value of $J$, the average density is finite, and we can
use mean field theory.   The mean field theory is
simple in this case, predicting  
\beq{32}
c[J(l^*), t = \infty; l^*] = [J(l^*) / \lambda(l^*)]^{1/2} \ .
\eeq
We match the mean field result to the observed value using the
scaling relation
\beq{33}
c(J) = e^{-d l^*} c[J(l^*); l^*] \ .
\eeq
We find for the concentration in the limit $J \to 0$:
\beq{34}
c(J) \sim \left\{
\begin{array}{ll}
\left( \frac{J}{\lambda} \right)^{1/2} & , d > 2 \\[0.1in]
\left[ \frac{J \ln(J_0/J)}{16 \pi D} \right]^{1/2} & , d = 2 \\[0.1 in]
\left( \frac{J_0 }{ \lambda^* }\right)^{1/2} 
\left( \frac{J}{J_0} \right)^{d / (2 + d)}
& , d < 2
\end{array} 
\right. ~\mathrm{as}~ J \to 0.
\eeq
Here, we have used the fact that $\lambda(l)$ goes to a fixed point
value for $d < 2$, just as it does in the case of no
adsorption.  These predictions are consistent with those of
Rey and Droz  \cite{Rey}.  Our argument that the long-range
nature of the reaction is irrelevant is, again, valid
when $d > 2$ or $d \le 2 $ and $n>2$, as shown in section \ref{sec4a}.

\subsection{The $A+B \to \emptyset$ Reaction}
\label{sec3b}

For the two-species reaction, we find the same flow equations
as for the single-species reaction.  We also find that the local
interaction dominates over the long-range component.
The matching, however,
is somewhat different.  Again, this difference is due to the
reactant segregation that can occur in low dimensions.
The appropriate mean-field equations in the matching limit are
\bey{35}
\frac{\partial c_A}{\partial t} &=& D \nabla^2 c_A - \lambda c_A c_B + J_A
\nonumber \\
\frac{\partial c_B}{\partial t} &=& D \nabla^2 c_B - \lambda c_A c_B + J_B 
\ ,
\eey
where $J_A$ and $J_B$ are random, Poisson adsorption fluxes of $A$ and
$B$ reactants, respectively. Note that 
$\langle J_A \rangle =
\langle J_B \rangle = J$.
  The initial conditions are
irrelevant in the long time limit, since the steady state
behavior is controlled by the adsorption rate.  We assume for
simplicity that $n_0 = 0$.
These equations can be simplified and
analyzed in a fashion similar to the case of no adsorption
\cite{Lee2,Deem2}.  Using the same variables as in section \ref{sec2b},
we find
\beq{36}
\frac{\partial \phi}{\partial t} = D \nabla^2 \phi + \delta J
\eeq
and
\beq{37}
\frac{\partial \rho}{\partial t} = D \nabla^2 \rho - 
\lambda [ \rho^2 - \phi^2] + \frac{J_A + J_B}{2}
\ ,
\eeq
where $\delta J = (J_A - J_B)/2$.
We can solve equation \ref{36} to find
\beq{37a}
\hat \phi(\vec{k}, t) = \int_0^t d t' e^{-D (t-t') k^2} \delta \hat 
J(\vec{k}, t') \ .
\eeq
 Averaging equation \ref{37}, we find
\beq{37b}
0 = -\lambda [ \langle \rho^2 \rangle  - \langle \phi^2 \rangle ] + J \ .
\eeq
From equation \ref{37a}, we find
\bey{38}
\langle \phi^2 \rangle &=& \lim_{t \to \infty} \int_\vec{k} 
\int_0^t d t' e^{- 2 D (t-t') k^2} \frac{J}{2}
\nonumber \\
&=& \lim_{t \to \infty}  \frac{J}{2} \int_0^t d t'\,
 \frac{e^{-4 d D t / h^2} }{h^2}
I_0^d \left( \frac{4 D t }{h^2} \right)
\nonumber \\
&=& \frac{J}{2 D} f_d  \ ,
\eey
where $f_d$ is a dimensionless
numerical constant depending on the dimension.
Here $I_0(x)$ is the modified Bessel function of the first kind.
For two or fewer dimensions, $f_d = \infty$.
We can bound the solution of equation \ref{37} \cite{Lee2}.
First, we note that since
the concentrations are non-negative,
$\rho   \ge  \vert \phi \vert$,
and so 
$\langle \rho^2 \rangle  \ge \langle \phi^2 \rangle$.
Using this, we find
\bey{39}
\langle \rho - \vert \phi \vert \rangle^2
 &\le&
\langle (\rho - \vert \phi \vert )^2 \rangle
\nonumber \\ 
&=&
\langle \rho^2 \rangle+
\langle \phi^2 \rangle - 2
\langle \rho \vert\phi\vert \rangle
\nonumber \\
&\le &
\langle \rho^2 \rangle-
\langle \phi^2 \rangle
\nonumber \\
&=& J/\lambda \ .
\eey
Combining the lower and upper bounds, we find
\beq{40}
\left[ \frac{J f_d}{\pi D} \right]^{1/2} \le 
\langle \rho \rangle
\le
\left[ \frac{J f_d}{2 D} +  \frac{J}{\lambda} \right]^{1/2} \ .
\eeq
We can immediately conclude that for a local reaction
\beq{40a}
\langle \rho \rangle = \infty,~~~ d \le 2 \ .
\eeq
This occurs because any given lattice site is dominated either
by $A$ or $B$ species due to segregation.  The random rate of
adsorption of the dominant species is so large for
$d \le 2$ that it swamps the relaxation by
diffusion of the species to nearest
neighbor lattice sites, which is the only mechanism for
reaction in a segregated system.
Since the local reaction is ineffective for $d \le 2$, we must
reconsider our argument that
the long-range component of the interaction
is irrelevant.  Indeed, the results of
section \ref{sec4b} imply that equation \ref{40a} holds only for
$n \ge 2 d$.

In the matching limit, for $d > z$, the flow equations lead to
\bey{41}
D(l) &=& D e^{(z-2) l}
\nonumber \\
\lambda(l) &=& \lambda e^{(z-d) l}
\nonumber \\
J(l) &=& J e^{(z+d) l} \ ,
\eey
where we set $z=2$ to obtain a fixed point for the
diffusivity.  We define $\epsilon = J(l) f_d / [2 D(l)]$ and
note that $\epsilon \ll J(l) / \lambda(l)$.  We see
that 
\beq{42}
\langle [ c_A(l) - c_B(l)]^2 \rangle = 4 \epsilon
\eeq
and
\beq{43} 
\langle [ c_A(l) + c_B(l)]^2 \rangle = \frac{4 J(l)}{\lambda(l)}
 + 4 \epsilon  \ .
\eeq
We note that
equation \ref{38} implies, since $\phi$ is a Gaussian field,
\beq{44}
\langle \vert c_A(l) - c_B(l) \vert \rangle = 
2 \left[\frac{2 \epsilon}{\pi} \right]^{1/2} \ .
\eeq 
This implies that $c_A(l)$ and $c_B(l)$ fluctuate in the
same way: $c_A(l) \sim c_B(l) + O(\sqrt \epsilon)$.  Since
$\langle \rho^2(l) \rangle = 
\langle \phi^2(l) \rangle  + J(l) / \lambda(l)$,
we find
$\langle c_A(l)^2 \rangle =
\langle c_B(l)^2 \rangle = J(l)/\lambda(l) + O(\epsilon)$.
Using
$\langle c_i(l)^2 \rangle =
\langle c_i(l) \rangle^2 + 
 \langle [\delta c_i(l)]^2 \rangle$ and
equations \ref{42} and \ref{43}, we find
$\langle [\delta c_A(l)]^2 \rangle =
\langle [\delta c_B(l)]^2 \rangle = O(\epsilon)$.
Since the fluctuations are much smaller than the mean, we 
can finally conclude that
\beq{45}
\langle c_A(l) \rangle
= \langle c_B(l) \rangle
\sim 
\left[ \frac{J(l)}{\lambda(l)} \right]^{1/2}~ \mathrm{as}~ J \to 0,~~~ d>2\ .
\eeq
Since the
scaling predicted by the flow equations
is trivial, this result holds for the observable concentrations as well:
\beq{45aa}
\langle c_A \rangle
= \langle c_B \rangle
 \sim \left\{
\begin{array}{ll}
 \left( \frac{J}{\lambda} \right)^{1/2}&, d > 2\\[0.1in]
\infty &, d \le 2
\end{array} \right.
 ~\mathrm{as}~ J \to 0.
\eeq

\section{Immobile Reactants with Adsorption}
\label{sec4}

We now consider the final case of immobile reactants
in the presence of adsorption.  We consider the
regime $J \to 0$, which leads to non-trivial scaling of
the reactant concentration with the adsorption rate.
As in section \ref{sec2}, we
will find that the decay rate
depends sensitively on the interaction $w(r)$.

\subsection{The $A+A \to \emptyset$ Reaction}
\label{sec4a}

For the single species reaction, we find the same
flow equations and dynamical exponent as in section \ref{sec2a}, with
the additional equation for the adsorption rate
\beq{45a}
\frac{d \ln J }{d l} = z+d \ .
\eeq
As in section \ref{sec2a}, we find that while simple scaling
arguments predict the local reaction
mechanism to dominate over the long-range reaction mechanism, this
is actually not the case.  In the low-concentration, small-$J$
limit, the reaction proceeds only by the long-range
mechanism for immobile reactants.  In the 
matching limit, we find $c[J(l^*); l^*] = [J(l^*)/J_0]^{1/2}$, where
$J_0 \approx \gamma h^{-d-n}$.  The observable concentration
is given by the scaling relation 
$c(J) = e^{- d l^*} c[J(l^*); l^*]$, where $l^*$ is chosen so
that $J(l^*) = J_0$.  Performing the matching, we find
\beq{46}
c(J) 
 \sim \left\{
\begin{array}{ll}
\left( \frac{J}{J_0} \right)^{d/(n+d)}  &, n>d\\[0.1in]
\infty &, n \le d
\end{array} \right.
 ~\mathrm{as}~ J \to 0.
\eeq
This result is the same as that found by Oshanin and coworkers
using an approximate, mean-field approach \cite{Oshanin2}.

We can also find an exact scaling relation for this problem.
Defining $\vec{x}^* = (J/\gamma)^{1/(n+d)} \vec{x}$
and $t^* = J (\gamma/J)^{d/(n+d)} t$, we find
\beq{47}
c(\vec{x}, t) = \left( \frac{J}{\gamma}\right)^{d/(n+d)}
c^*(\vec{x}^*, t^*) \ ,
\eeq
where $c^*$ satisfies the exact master equation \ref{1} with
$J = 1$ and $\gamma = 1$.  This scaling relation implies
that $\langle c \rangle = (\mathrm{const}) (J/\gamma)^{d/(n+d)} =
(\mathrm{const}) (J/J_0)^{d/(n+d)}$, which is exactly what we
found in equation \ref{46}.

\subsection{The $A+B \to \emptyset$ Reaction}
\label{sec4b}

For the two-species reaction, we find the same flow equations
and dynamical exponent as in section \ref{sec2b}, with
the addition of equation \ref{45a}.  In the
matching limit, we can use a similar type of mean
field theory:
\beq{48}
\frac{\partial \phi}{\partial t} = \rho (\phi * w)
-\phi \rho \hat w(0) + \delta J
\eeq
and
\beq{49}
\frac{\partial \rho}{\partial t} = \phi (\phi * w) -
\rho^2 \hat w(0)   + (J_A + J_B)/2 \ .
\eeq
Equation \ref{48} can be solved, since $\rho$ depends on time only:
\bey{50}
\hat \phi(\vec{k}, t ) &= &
\int_0^t d t'
\exp\left\{ - \int_{t'}^t d t''[\hat w(0) - \hat w(k)]
\rho(t'') \right\}
\nonumber \\ && \times
 \delta \hat J(\vec{k}, t') \ .
\eey
Taking the average of equation \ref{49} and using
equation \ref{50} and $\langle \delta \hat J(\vec{k}, t')
\delta \hat J(\vec{k}', t'') \rangle = (J/2) \delta(t'-t'')
(2 \pi)^d \delta(\vec{k} - \vec{k}')$, we find
\bey{51}
\frac{d \langle \rho \rangle}{d t} &=& \frac{J}{2}
\int_0^t d t' \int_\vec{k} 
\nonumber \\ && \times \hat w(\vec{k}) \left\langle \exp\left\{
-2\int_{t'}^t d t'' [\hat w(0) - \hat w(k)]
\rho(t'')
\right\} \right\rangle
\nonumber \\
&&- \left\langle \rho^2 \right\rangle \hat w(0)  + J \ .
\eey
In the long-time limit, the average density goes to a constant.
In this limit, then,
\beq{52}
0 = \frac{J}{2} \int_\vec{k} \frac{\hat w(\vec{k})}{2 [ \hat w(0) -
\hat w(\vec{k})] \rho} - \rho^2 \hat w (0) + J \ .
\eeq
Using the explicit expression \ref{8} for the interaction, we
can evaluate the first integral:
\bey{52a}
\int_\vec{k}  \frac{\hat w(\vec{k})} {\hat w(0) - \hat w(k)}
&=& (\mathrm{const}) \int_0^\Lambda d k\, \hat w(k) k^{d-\zeta - 1} \ , 
\nonumber \\
&<& \infty,~~~~~~ d > \zeta \ .
\eey
Here $\Lambda = 2 \pi/h$ is the cutoff in Fourier space, and
we have concentrated on the small-$k$, large-distance properties
of this integral.  Since $\zeta = n-d$, we see that
the integral converges to a finite value
only for $n < 2 d$.
When the integral converges,
 the first term in equation \ref{52} is subdominant, 
because $\rho$ is large in the matching limit, and by performing the matching
we find the expected mean-field behavior.  In the case of $n \ge 2 d$, 
the average density diverges, as we can see explicitly from the
bound $\langle \rho \rangle \ge \langle \vert \phi \vert \rangle$, since
$\langle \phi^2 \rangle = (\mathrm{const}) \int_0^\Lambda d k\,
k^{d-\zeta-1}$ diverges for $n \ge 2 d$.
The final result is
\beq{53}
c_A = c_B \sim
\left\{
 \begin{array}{ll}
\infty & , n \ge 2 d\\[0.1in]
\left(\frac{J}{J_0}\right)^{d/(n+d)} & , d < n < 2 d\\[0.1in]
0 & , n \le d
\end{array}
\right. ~\mathrm{as}~ J \to \ 0.
\eeq
This result satisfies the exact scaling relation  \ref{47}, as it must.

Interestingly, if the interaction decays too quickly with 
distance, $n \ge 2 d$, then
segregation occurs, and the adsorption swamps the reaction rate.
This effect can only occur in the $A+B \to 0$ case, where the
density of a majority species can grow without bound on a lattice site.
If the interaction is sufficiently long-ranged, $d < n < 2 d$, then
segregation is prevented, and the reaction proceeds as in the
homogeneous, single-species case.  As always, if the interaction
is not integrable, $ n \le d$, the concentration is forced to
vanish in the limit of a large system.

\section{Summary}
\label{conclusion}

Using a field-theoretic formulation of reaction kinetics, we have
derived the asymptotic behavior of long-ranged, bimolecular reactions.
When the reactants are mobile, the long-ranged nature of the
reaction is usually irrelevant.  Conversely, when the reactants
are immobile, the long-ranged reactive interaction is crucial for the
reaction to proceed.  In the $A+A \to \emptyset$ reaction, concentration
fluctuations are only mildly important.  The reactants in 
the $A+B \to \emptyset$
reaction, however, typically segregate in the absence of 
adsorption.  Adsorption tends either to homogenize or to drive the system
to infinite density in the $A+B \to \emptyset$ reaction.

\begin{acknowledgement}
This research was supported by the National Science Foundation
through grant CTS--9702403.
\end{acknowledgement}

\bibliography{react5}

\end{document}